\documentclass[twoside,leqno,singlecolumn,letter]{article}
\usepackage{ltexpprt,url,amsmath,amsfonts,amssymb,graphics,graphicx,color}
\usepackage[utf8]{inputenc}
\usepackage{authblk}
\usepackage[labelfont=bf,labelsep=newline]{caption}
\providecommand{\keywords}[1]{\textbf{\textit{Keywords---}} #1}

\title{\bf{Cell Identity Codes: Understanding Cell Identity from Gene Expression Profiles \\using Deep Neural Networks}}

\author[1]{Farzad Abdolhosseini}
\author[2]{Behrooz Azarkhalili}
\author[1]{Abbas Maazallahi}
\author[1]{Aryan Kamal}
\author[1]{\\Seyed Abolfazl Motahari}
\author[1]{Ali Sharifi-Zarchi*}
\author[3]{Hamidreza Chitsaz*}
\affil[1]{Department of Computer Engineering, Sharif University of Technology, Tehran, Iran}
\affil[2]{Royan Institute for Stem Cell Biology and Technology, ACECR, Tehran, Iran}
\affil[3]{Department of Computer Science, Colorado State University, Fort Collins, CO, USA}
\affil[*]{Corresponding authors: asharifi@sharif.ir, chitsaz@chitsazlab.org}

\date{}
\begin{document}
\maketitle

\begin{abstract}\small%\baselineskip=9pt

Understanding cell identity is an important task in many biomedical areas. Expression patterns of specific marker genes have been used to characterize some limited cell types, but exclusive markers are not available for many cell types. 
A second approach is to use machine learning to discriminate cell types based on the whole gene expression profiles (GEPs). The accuracies of simple classification algorithms such as linear discriminators or support vector machines are limited due to the complexity of biological systems. 

We used deep neural networks to analyze 1040 GEPs from 16 different human tissues and cell types. After comparing different architectures, we identified a specific structure of deep autoencoders that can encode a GEP into a vector of 30 numeric values, which we call the {\em cell identity code} (CIC). The original GEP can be reproduced from the CIC with an accuracy comparable to technical replicates of the same experiment. Although we use an unsupervised approach to train the autoencoder, we show different values of the CIC are connected to different biological aspects of the cell, such as different pathways or biological processes. This network can use CIC to reproduce the GEP of the cell types it has never seen during the training. It also can resist some noise in the measurement of the GEP.
Furthermore, we introduce {\em classifier autoencoder}, an architecture that can accurately identify cell type based on the GEP or the CIC. 

\keywords{
Deep learning, autoencoder, gene expression profile, artificial neural network, classifier autoencoder.
}

\end{abstract}

\section{Introduction}
Accurate identification of cell types has received significant attention due to several applications in research and clinics. For instance, one major goal of regenerative medicine is to differentiate pluripotent stem cells, such as embryonic stem cells (ESCs) or induced pluripotent stem cells (iPSCs), to  some specific tissue required for a patient, such as neurons or cardiomyocytes \cite{Anonymous:eV193zNk}. After treating the stem cells with a differentiation protocol, an important question is whether the produced cells have obtained  true identity of the desired cell type \cite{Volarevic:2011ke}. 

There are some specific marker genes identified whose expression patterns can be used to characterize different cell types. Some examples are the expression of Insulin, which confirms the $\beta$-cell identity;  pluripotency markers such as Nanog which are used to identify pluripotent stem cells; and membrane markers that label several types of cancer cells. The application of marker genes, however, is limited to specific cell types. For many cell types, no markers are identified so far that can accurately characterize them. Furthermore, the expression of many marker genes is not exclusive to one particular cell type. For instance, Oct4 that is used in many studies as a pluripotency marker \cite{Leitch:2013gz} is also expressed in non-pluripotent cell types such as adult stem cells \cite{Liedtke:2008cb}. One important possible solution is to recognize the cell identity using the whole-genome gene expression profile (GEP), rather than a small subset of marker genes \cite{Kuo:1992hq,Vidarsson:2010fv}.  

Despite its importance, there have been limited efforts to systematically recognize cell identity through the GEP. Classical differential expression analysis and clustering methods are used to discriminate a limited number of cells. An extension of the nearest centroid classification method was suggested for classification of cancer types \cite{Tibshirani:2002ht}. This method and empirical Bayes analysis \cite{Efron:2011jj} were used to classify 88 normal and prostate cancer samples into different sub-types \cite{Stuart:2004bx}. Linear algebraic and statistical methods such as non-negative matrix factorization, Kullback-Leibler divergence and non-negative least squares were employed in an unsupervised method for cell type identification. We applied it to 77 samples for discriminating three cell types \cite{Zuckerman:2013hx}. Although shown to be successful for a few cell types, whether these methods can be extended to analyze tens of different cell types needs further investigation. 

Although there are many models developed so far to classify a few different cell types, there are few generic models to classify an arbitrarily large number of cell types with good accuracy, one of which is CellNet \cite{Cahan:2014fh}. It is an online tool that can analyze Microarray gene expression profiles of mouse or human and assign them to each of 16 human or 20 mouse tissues or cell types. CellNet requires a minimum 60 samples of each tissue to derive the global and tissue-specific gene regulatory networks. The subnetworks of the union of these networks are then identified using a community detection algorithm called Infomap \cite{Bohlin:2014fr}, from which the tissue-specific subnetworks are identified using Gene Set Enrichment Analysis (GSEA) \cite{Subramanian:2005jt}. Expression levels of the genes belonging to the tissue-specific subnetworks are then used to train Random Forest binary classifiers for each tissue. As a limitation, 
it is a cumbersome task to hand-pick an equal number of samples from each cell types that have the same Microarray platform and undergo perturbations, in order to derive the gene regulatory networks. Furthermore, deriving a highly accurate gene regulatory network by only using GEPs is challenging.

A major obstacle to classical machine learning methods such as Support Vector Machine (SVM) or Random Forest is the presence of hundreds, or even thousands of genes,  which are differentially expressed among different cell types, without causal connection to the cell identity. Their expression change can be even more significant than the master regulators of the cell identity due to a number of reasons, such as differences among batches, culture conditions, strains, or treatment protocols of different cell types. As a result, a simple classifier might highlight a group of genes as discriminators of the cell identity for one training dataset, which fail to correctly predict the identities of another dataset. 

Recently, deep learning has made breakthroughs in different scientific areas such as games \cite{Mnih:2015jp, Silver:2016hl}, speech \cite{Deng:gk}, face \cite{Sun:2014uv}, image and text \cite{Srivastava:2012vd} recognition, robotics \cite{Lenz:2015ih}, and web search \cite{Huang:2013fz}. It has also been  used in several bioinformatics applications including predicting protein binding sites in DNA and RNA \cite{Alipanahi:2015fb}, DNA replication initiation and termination zones \cite{Liu:2015iq}, protein secondary structure \cite{Heffernan:2015gh} and folding \cite{Jo:2015el}, residue-residue and protein-protein interaction \cite{Du:2016dn}, non-coding DNA  function prediction \cite{Quang:2016jt} and inferring expression of target from landmark genes \cite{Chen:2016jba}. 

Deep autoencoders are a group of deep neural networks that were originally developed to learn dimensionality reduction. They are usually given some data as the input, and they are expected to generate the same data as the output. The data needs to pass an internal layer that has significantly fewer neurons than the dimensionality of the data, hence it learns to encode and decode the data with the minimum loss. They have received considerable attention due to several reasons. They are usually unsupervised and can be trained using unlabelled data. They can learn internal patterns of the data, such as correlations among different variables, in order to reduce its dimensionality. Furthermore, there are several extensions of deep autoencoders that can perform additional tasks such as denoising the data \cite{Vincent:uo}.

Due to the internal patterns of the GEPs such as coexpression of large clusters of genes we hypothesized the dimensionality of GEPs can be significantly reduced using deep autoencoders. By doing so, we can transform the GEP into a smaller number of features that can reproduce the whole GEP. We also hypothesized that these features might reflect different aspects of the cell biology. Furthermore, we speculated the cellular identity can be revealed by these features. Here we employ and extend deep autoencoders to address these hypotheses.

\section{Results and Discussion}

We obtained normalized expression profiles of 20184 genes in 1040 samples of 16 different human tissues and cell types from different datasets of NCBI GEO, which were collected and preprocessed in CellNet  \cite{Cahan:2014fh}. There was an equal number of 65 samples per cell type or tissue with at least 10 different biological perturbations to ensure the diversity of the samples from each tissue. Presence of different datasets and perturbations was important to prevent overfitting.

\begin{figure}[h!]
	\begin{center}
		\includegraphics[width=\textwidth]{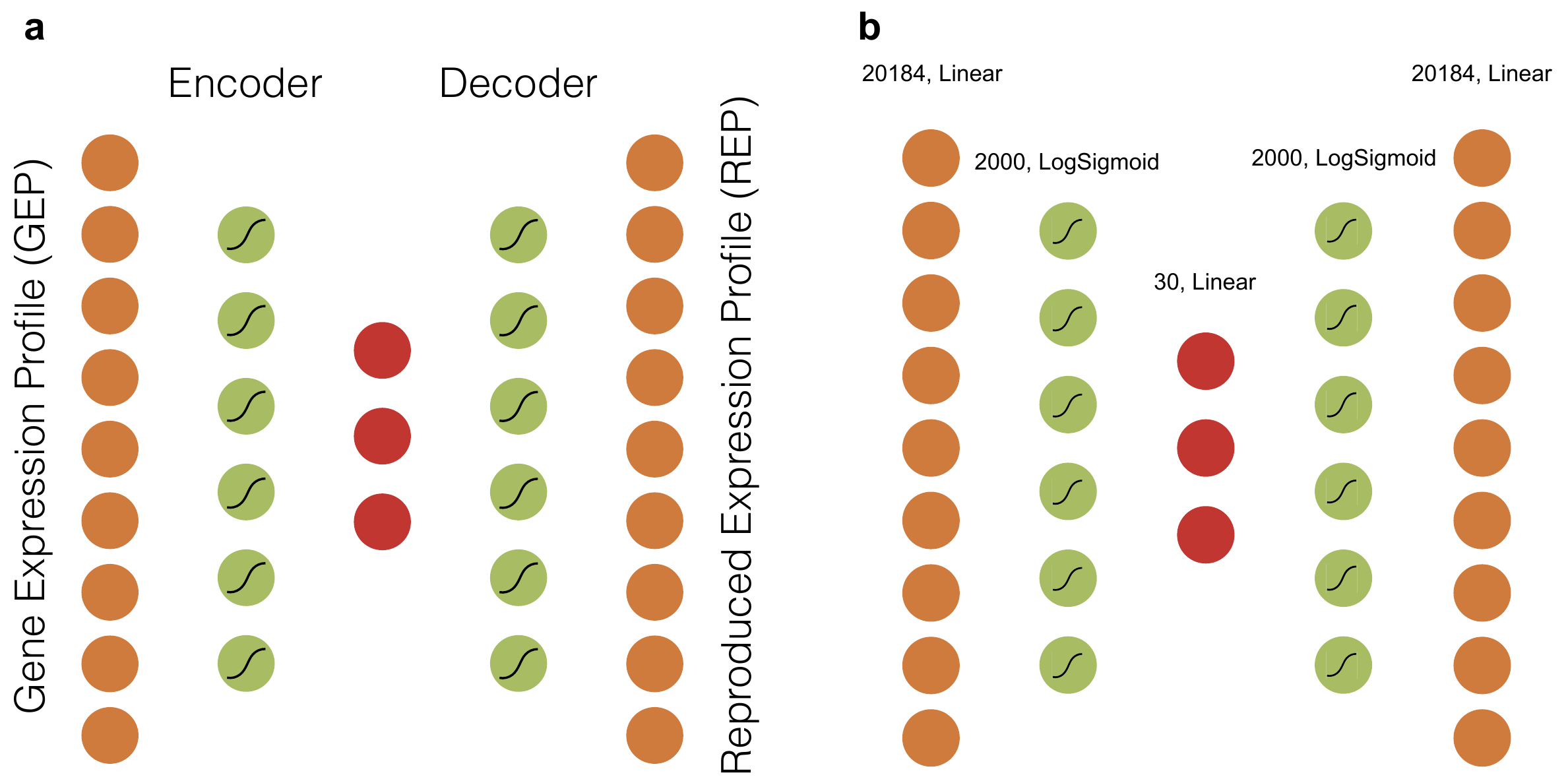}
	\end{center}
	\caption{}
	\label{fig:ae} 
\end{figure}

Fig. \ref{fig:ae}a shows a general architecture of an autoencoder, consisting of an encoder part that converts a given gene expression profile (GEP) to a code layer shown in red, and then decodes it back to reproduce the original GEP. Hereafter we call the output of the encoder part the Cell Identity Code (CIC), and the output of the decoder part as the Reproduced Expression Profile (REP). 

Our first task was to determine a particular architecture of autoencoders that can accurately reproduce GEPs. For this purpose, we tested 10 different architectures with distinct layer configurations and activation functions. After selecting the outperforming architecture, we tested different sizes of the CIC layer and selected a particular architecture with 30 neurons in the CIC layer, as shown in Fig. \ref{fig:ae}b. See methods for more details. The results of the comparison between different neural network architectures and between different sizes of CICs are presented in subsections \ref{sec:architectures} and \ref{sec:dimensions}.

\subsection{Accuracy of Cell Identity Codes}

We asked whether a 20184-dimensional GEP vector can be accurately compressed in a 30-dimensional CIC, without losing the data. To address this question, we trained the selected network architecture (Fig. \ref{fig:ae}b) by 75\% of randomly selected samples ($n=780$) as the training dataset. Then we measured its performance on the remaining 25\% of the GEPs ($n=260$), as the test dataset. Each test sample was encoded to a CIC and then decoded to a reproduced gene expression profile (REP). The distance between each REP and the original GEP was measured by Mean Square Error (MSE). We observed a stable decreasing MSE trend during the training. 

\begin{figure}[h!]
	\begin{center}
		\includegraphics[width=.7\textwidth]{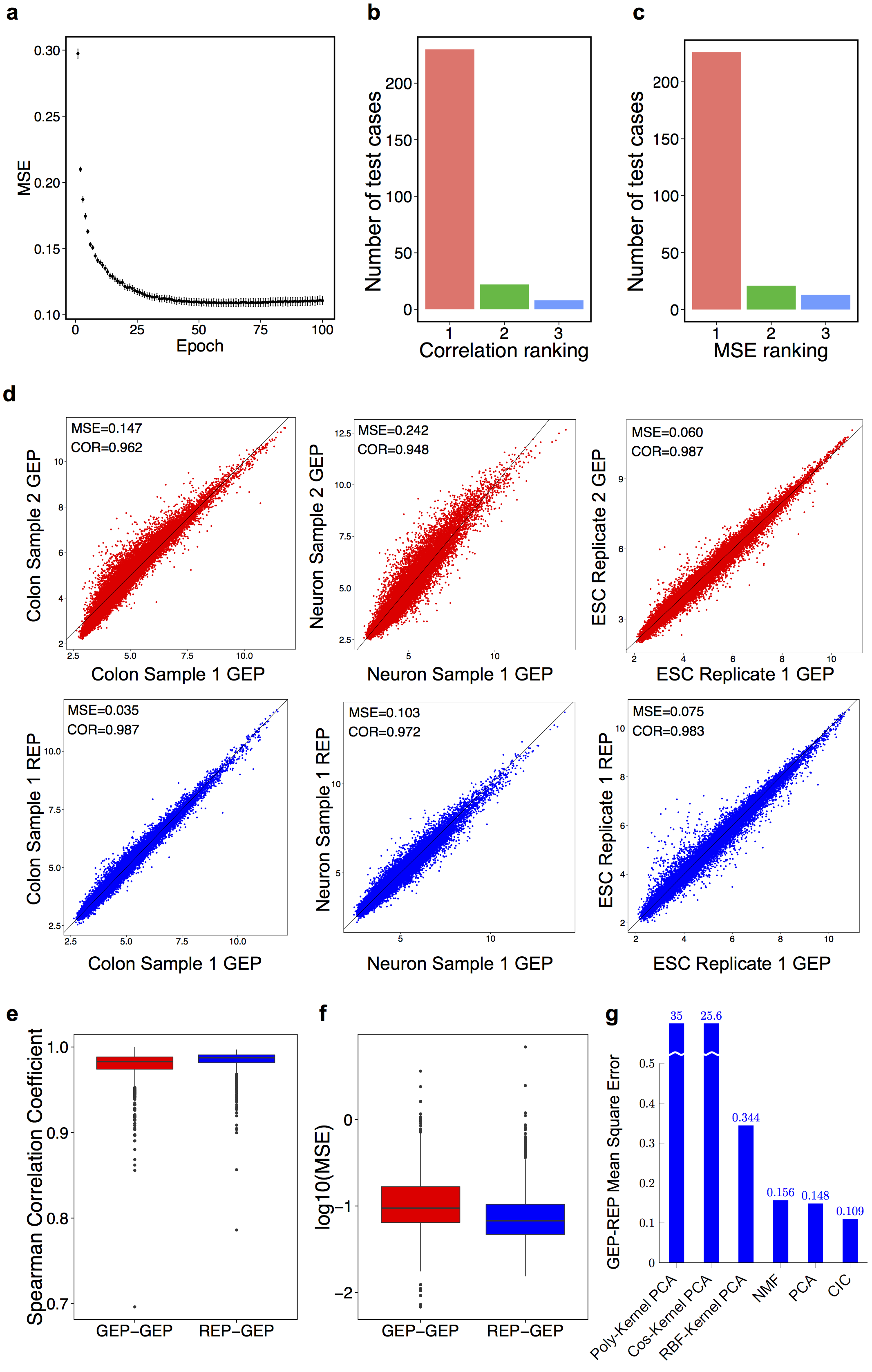}
	\end{center}
	\caption{}
	\label{fig:allgenes}
\end{figure}

To ensure the robustness of the results and lack of overfitting, we also performed a 10-fold cross-validation with balanced sampling.  As shown in Fig \ref{fig:allgenes}a, the average MSE of the test dataset is significantly decreased during the training from 0.3 (epoch 1) to 0.11 (epoch 66). The final value of the average MSE at epoch 100 is also 0.11, which shows 100 training epochs was sufficient. The error bars depict the standard errors of the measured MSEs among  cross-validation runs. The maximum standard error during the training is 0.0035, which is much smaller than the average MSEs (maximum ratio < 3\%), which shows the measured average MSEs are robust, and there is no sign of overfitting to a particular portion of the data.

While the average MSEs showed a small average distance between each pair of GEP-REP, we questioned whether this holds for each individual pair. To answer that, for each GEP in the test data, we counted the number of other GEPs that were closer than the paired REP. There was an average number of about 16 samples per tissue in the test data, many of which were from biological or technical replicates. Hence if a REP was too different from its GEP, it was likely that many other test samples were closer to the GEP. Fig. \ref{fig:allgenes}b shows the results, in which the Spearman correlation coefficient is used as the measure of similarity between a pair of GEP-REP, or between two GEPs. For 88.4\% of the test cases (230 out of 260), the paired REP was ranked the first, which means the REP has been closer than any other test sample. For 22 cases (8.4\%) the REP was ranked the second, and for 8 test cases (3\%) it was ranked the third. We also used MSE as a measure of distance and counted the number of test-cases for which the REP was the least distant than any other test sample (Fig. \ref{fig:allgenes}c).  For 226 test cases (87\%) the REP was ranked the first, for 21 cases (8\%) it was ranked the second, and for 13 cases (5\%) it was ranked the third. This showed the results are similar, regardless of the similarity or distance metric. Most of the samples for which the REP was ranked the third were embryonic stem cells (ESCs).

We also performed the reverse experiment: for each test-case REP, we sorted all test-case GEPs according to their similarity (measured by either higher correlation or lower MSE) and identified the ranking of the original GEP that matched the REP. For 100\% of the test-cases, the correct GEP was ranked the first. It means for each REP, its original GEP is more similar than any other test sample.

To further illustrate this, three examples are provided in Fig. \ref{fig:allgenes}d. In each column, the red scatter plot compares two GEPs denoted as samples 1 and 2, and the blue scatter plot compares the same sample 1 GEP with its REP. The colon and neuron samples (the left and middle columns) are carefully selected from the same study with an exactly matching region of the body, from two different human subjects. In both examples, the GEP of sample 1 is closer to its REP, than another sample from the same type and study. This is shown as higher correlation and lower MSE in blue scatter plots than the red ones. The ESC samples in the right column are replicates of the same cell line in the same study. There is a negligible distance between the MSE and correlation values of the blue versus the red scatter plots. It is important to note that ESCs are not primary tissues, rather they are {\em in-vitro} isogenic cell lines that are cultured in equal culture conditions. 

To further scrutinize this, we identified for each of $n=260$ test-cases the closest other GEP based on Spearman correlation coefficient (Fig. \ref{fig:allgenes}e) and MSE (Fig. \ref{fig:allgenes}f). Then we measured the similarity (correlation) or the distance (MSE) between the paired GEPs, and also between each GEP and its REP. As shown, the Spearman correlation values are significantly higher for REP-GEP than the GEP-GEP pairs (Wilcoxon {\em p}-value $<10^{-15}$). Also, the MSE values are significantly lower between REP-GEP pairs rather than GEP-GEP (Wilcoxon {\em p}-value $<10^{-15}$). 

Taken together, these experiments show the expression profiles are reproduced from the CICs with an accuracy comparable to profiling a technical replicate of the same sample.

Next, we compared the performance of our method against other widely-used dimensionality reduction algorithms. For that purpose, we transformed GEPs to 30-dimensional spaces using ordinary PCA, Non-negative Matrix Factorization (NMF), Polynomial-kernel PCA, Cosine-kernel PCA and Radial Basis Function (RBF)-kernel PCA. For each method, we used the inverse transformation to reproduce 20184-dimensional REP vectors and measured the MSE between each pair of GEP-REP. As shown in Fig. \ref{fig:allgenes}g, our CIC vectors outperform the other dimensionality reduction algorithms.

\subsection{Universality of Cell Identity Codes}

The next question was whether the autoencoder that is trained to generate the CICs based on some specific training cell types can also generate accurate cell identity for the other cell types, i.e. it can generalize to unseen cell types without retraining.

To answer this question we reserved all 65 B-cell samples for test and used 975 samples of all other 15 tissue types to train. After training an autoencoder for 30 epochs of 100 iterations, we generated cell identity codes (CICs) for the test cases, and subsequently reproduced the expression profiles of B-cells using CICs. In all 65 samples, the reproduced gene expression profiles were closer to the original gene expression profiles of each B-cell sample than to any of the other 975 gene expression profiles of the training dataset. Both correlation and MSE results were significantly better for REPs than for the samples of other tissues.

This observation was surprising since there were other groups of blood cell types such as T-cells or macrophages available in the dataset. If the autoencoder had not generalized, it would have produced cell identity codes for B-cells that were similar to the cell identity codes of T-cells or macrophages. As a result, the reproduced gene expression profiles of those cell identity codes would be closer to the GEP of T-cells or macrophages available in the training dataset, rather than those of the original B-cell sample. 

We also questioned whether the REPs have a B-cell biological identity. For that purpose, we performed a differential expression analysis between B-cell reproduced gene expression profiles and T-cell original gene expression profiles. We selected the genes that were significantly enriched ($\log_2$ fold-change $> 1$ and Benjamini-Hochberg adjusted {\em p}-value $\leq 0.05$) in B-cell REPs. This analysis identified 380  of such unique genes, that were used for a pathway analysis using Enrichr \cite{Enrichrinteractive:2013dv}. B-cell receptor signaling pathway of Homo sapiens was ranked the highest according to a Z-score criterion, with an adjusted {\em p}-value of 0.026.

This was a striking observation since the B-cell samples were never used in training and the autoencoder did not have any record of the B-cell gene expression patterns. When very well discriminated from the T-cells, we were confident we could obtain the same or even better results for comparison of the B-cell REPs with the GEPs of the other training cells. 

\begin{figure}[h!]
	\begin{center}
		\includegraphics[width=1\textwidth]{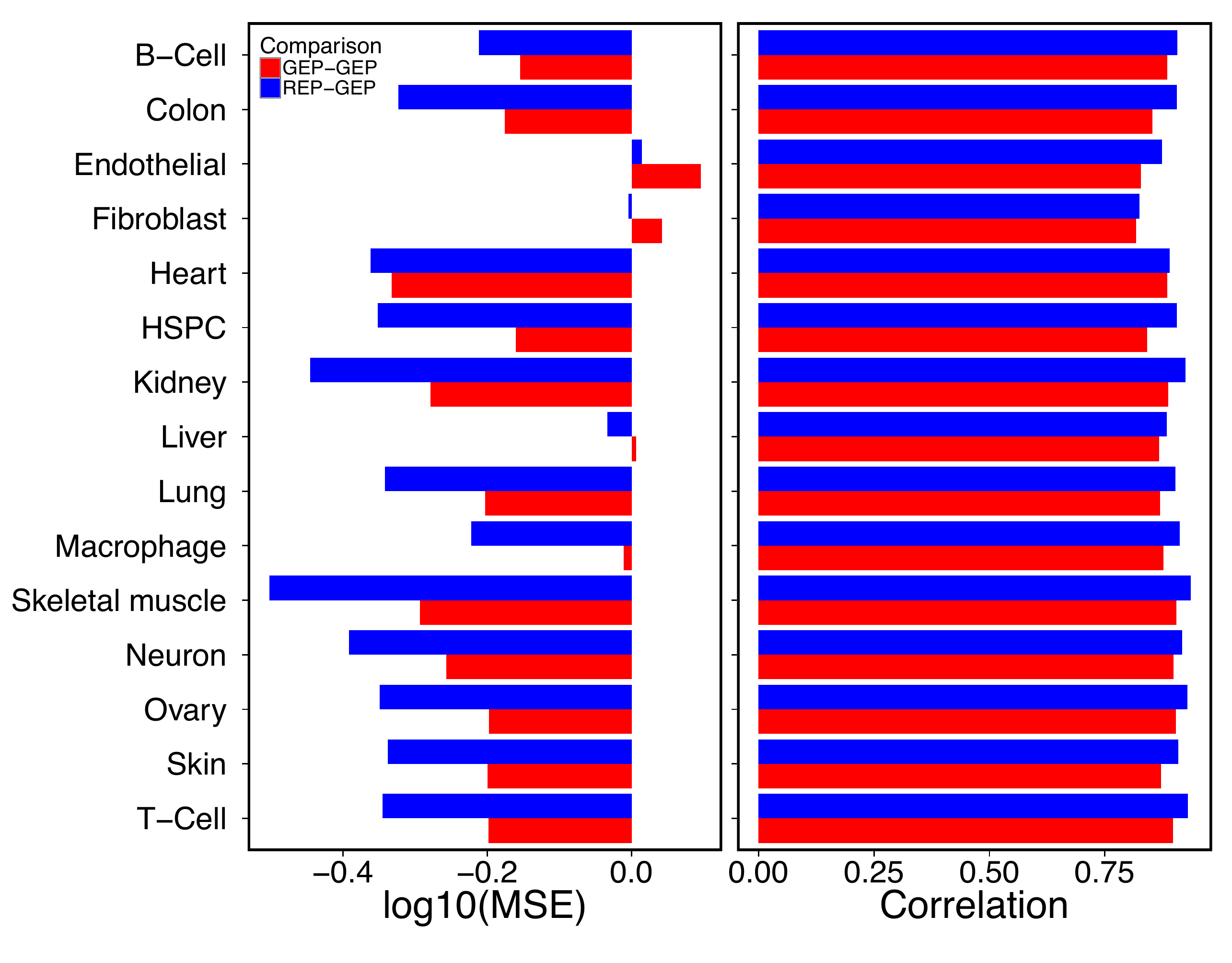}
	\end{center} 
	\caption{}
	\label{fig:generalize}
\end{figure}

To ensure this observation was not limited to the B-cells, we performed the same experiment for each of the primary tissue samples available in our dataset; see Fig. \ref{fig:generalize}. For each cell type, we excluded all of its samples from the training and tested the autoencoder with those samples. Although we trained autoencoders in a few epochs (less than 3 epochs for all tissues excepting B-cell and heart), both higher Correlation medians and lower MSE medians in almost all tissues show the expression profiles can be produced from cell identity codes with high accuracy even if the target tissue is not used in training of the algorithm.

\subsection{Cell Type Classification}

We then asked if cell identity codes can be used for characterization of the samples. For this purpose, we used the CICs produced by 100 epochs of training on 75\% of all samples (training for 200 epochs resulted in overfitting, see supplementary Fig. S8). Then we used two widely used classification algorithms Random Forest (RF) and Support Vector Machines (SVM) to determine the tissue or type of the cells, from the 16 available tissue/cell types, using cell identity codes. Both RF and SVM were trained on the same 75\% training samples that were used for training the autoencoder. 

To compare the accuracy of the results, we also used Principal Component Analysis (PCA) to reduce the dimension of the original gene expression profiles from 20184 to 30 dimensions, equal to the dimension of the cell identity codes. Both classification algorithms were also separately trained on the PCA transformation of the same training set. Then, all trained classifiers were applied to the test samples.

Out of 260 test cases, 25 and 27 samples were misclassified on the PCA transformation of the GEPs by Random Forest and SVM, respectively. Using cell identity codes, the number of misclassifications by Random Forest was slightly reduced to 16. SVM produced the same number of errors (supplementary Fig. S7).

Remarkably, there could be an unlimited number of CICs generated from the same gene expression profile depending on the parameters and weights of the neural network, with comparable accuracies of reproducing gene expression profile. All of those cell identity codes, however, would not be necessarily suitable for classification of cell types. We expected some CICs to have particular patterns for each cell type, while other CICs of the same sample could not be easily classified. 

\begin{figure}[h!]
	\begin{center}
		\includegraphics[width=0.5\textwidth]{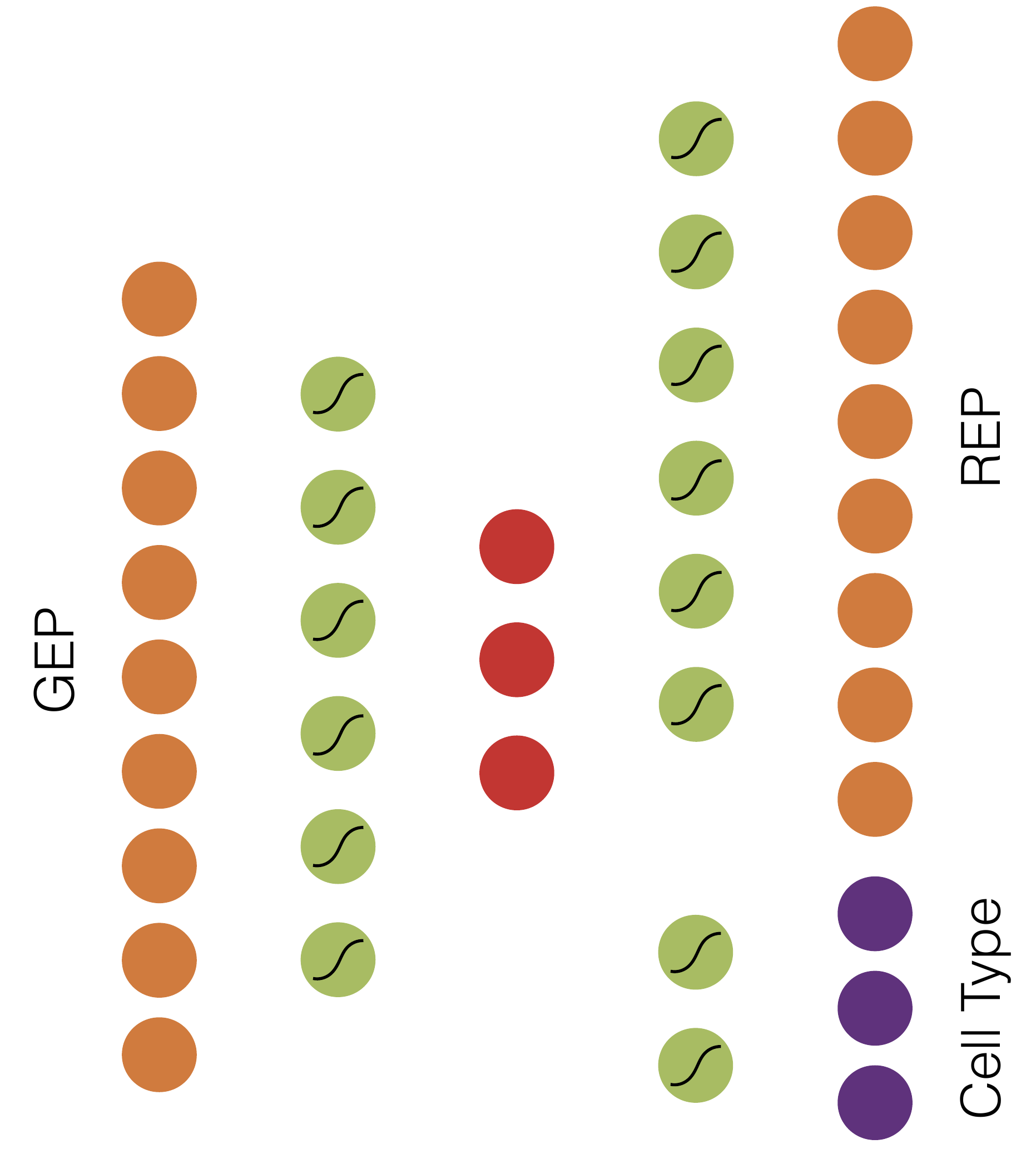}
	\end{center}
	\caption{}
	\label{fig:classifier}
\end{figure}

In order to guide the training process of autoencoder to identify those neural network parameters that could produce more easily classifiable cell identity codes, we incorporated some additional layers to the network architecture; see Fig. \ref{fig:classifier}. The new architecture, which we call classifier autoencoder, consists of two encoder and output sequential subnetworks, which were connected through the 30 neurons in the middle - i.e. the cell identity code neurons. The output consists of two branches, a decoder and a classifier. The classifier contained at most one hidden layer followed by a non-linear SoftMax layer. Therefore, the final code should also have the capacity to predict cell type by using a simple model (linear or neural network with only one hidden layer). Training was performed using a weighted average of two criteria functions, an MSE criterion for the decoder and a cross-entropy criterion for the classifier output.

This new architecture was used in the same way that we trained the simple autoencoder on the training set. Its performance on the test set outperformed both PCA and simple autoenoder. There were 10 misclassifications (3.8\% error) by both SVM and Random Forest methods. The 60\% error improvement in classifier autoencoder, in comparison with PCA, was an evidence for the capabilities of cell identity codes in determining the cell identity (supplementary Fig. S7).

\subsection{CIC Represents Aspects of Cell Biology}

An important question is whether biological pathways, processes, and other important aspects of cell biology can be learned from the expression data using cell identity codes. In other words, does each CIC component represent a different part of the cellular machinery? By CIC component, we mean each of the 30 numeric values in a cell identity code vector.

To answer this question, we sought to determine which subset of genes are most influenced by changing the $i$-th component of CIC ($1 \leq i \leq 30$). As an example, if the first CIC component represents the cell division process, then it should have the maximum effect on the reproduced expression values of the genes that are involved in cell division.

Let $C_i(x)$ be the $i$-th component of CIC for a given GEP $x$. We call $\bar{C}_i$ and $\sigma_i$ as the mean and standard deviation of $C_i(x)$ for all available samples $x$, respectively. We used the trained classifier autoencoder to compute the values $\bar{C}_i$ and $\sigma_i$  for all values $i (1 \leq i \leq 30)$. Then we produced a REP by feeding the $\bar{C}=(\bar{C}_1,\bar{C}_2,\cdots, \bar{C}_{30})$ as the input of the decoder network. We call the output of the decoder network for this particular input as the baseline REP. We also fed the decoder network 30 additional inputs with a value of $(\bar{C}_1,\bar{C}_2,\cdots,\bar{C}_{i-1},\bar{C}_i+2\sigma_i,\bar{C}_{i+1},\cdots,\bar{C}_{30})$. It means that for round $i$, we only changed the $i$-th component of $\bar{C}$ by adding $2\sigma_i$ to $\bar{C}_i$. Then we compared the output of decoder network to see the reproduced expression values of which genes are most increased, in comparison with the baseline REP. For each CIC component, we determined 100 genes with the highest absolute change in REP after increasing value of the component. We took an equal number of genes for each CIC component to prevent bias in statistical analyses. There were a few cases, that the same gene was present in sets of two or more CIC components.

\begin{figure}[p!]
	\begin{center}
		\includegraphics[width=0.9\textwidth]{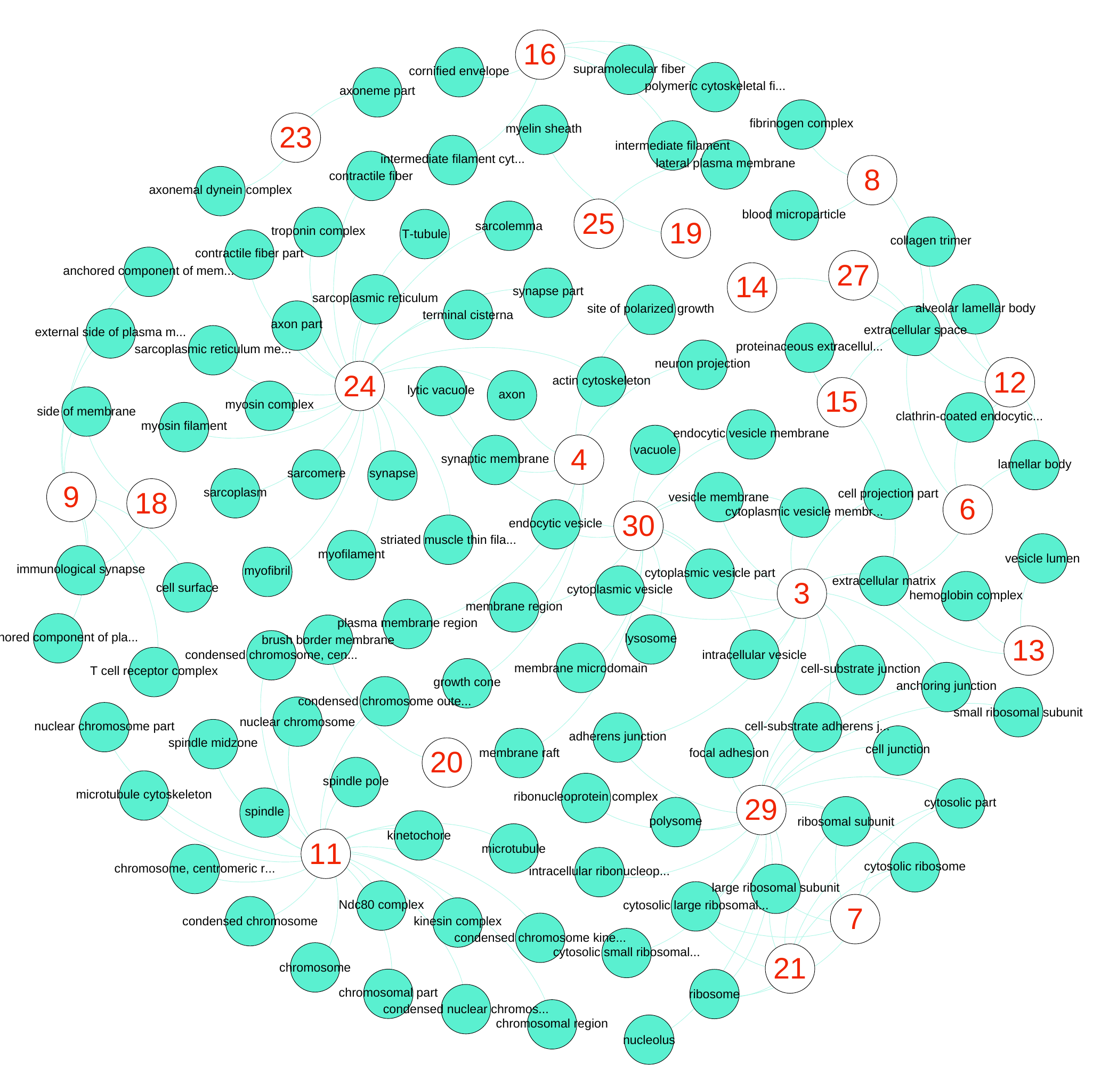}
	\end{center}
	\caption{}
	\label{fig:cc}
\end{figure}

\begin{figure}[p!]
	\begin{center}
		\includegraphics[width=0.9\textwidth]{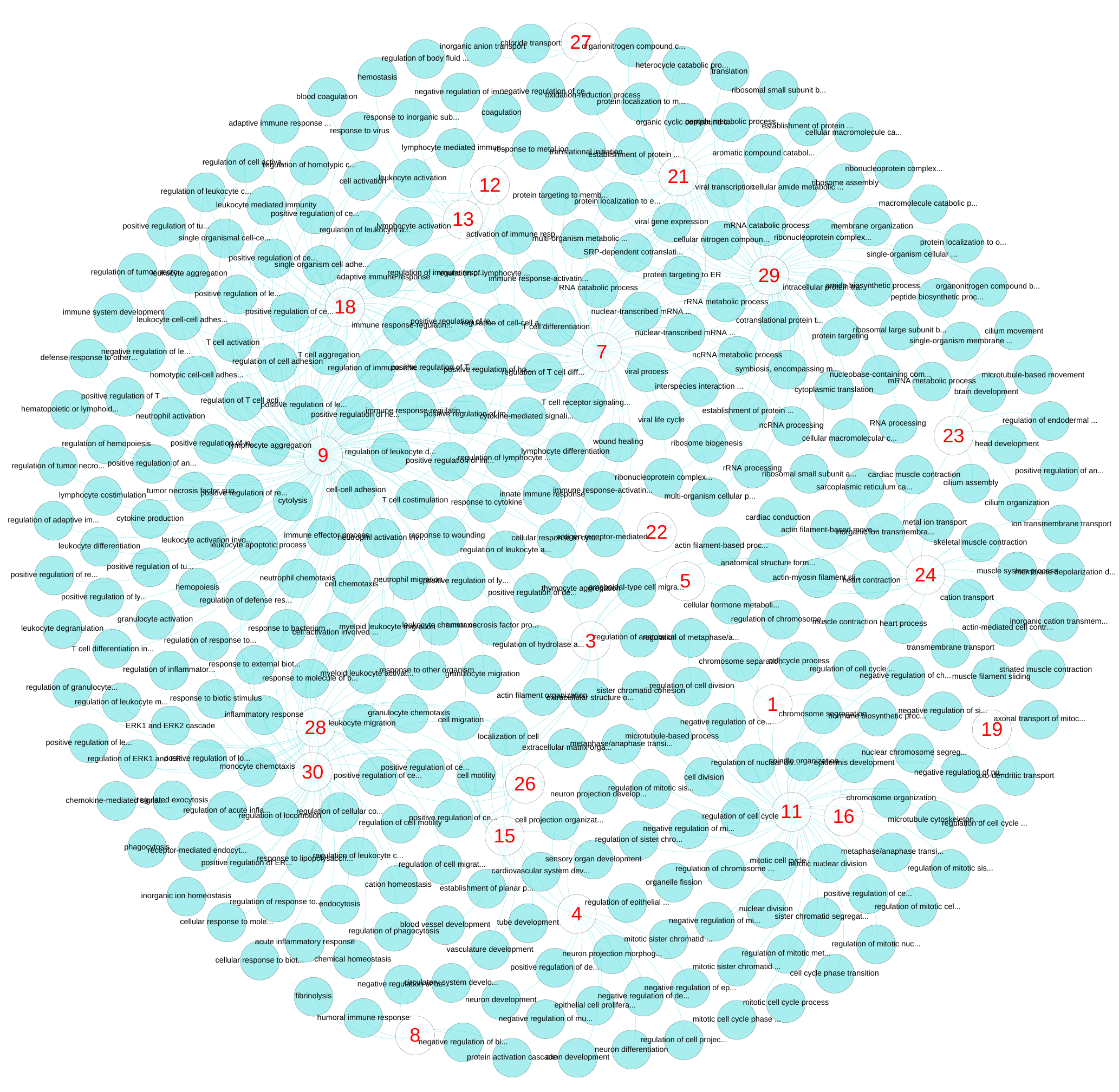}
	\end{center}
	\caption{}
	\label{fig:bp}
\end{figure}

\begin{figure}[p!]
	\begin{center}
		\includegraphics[width=0.9\textwidth]{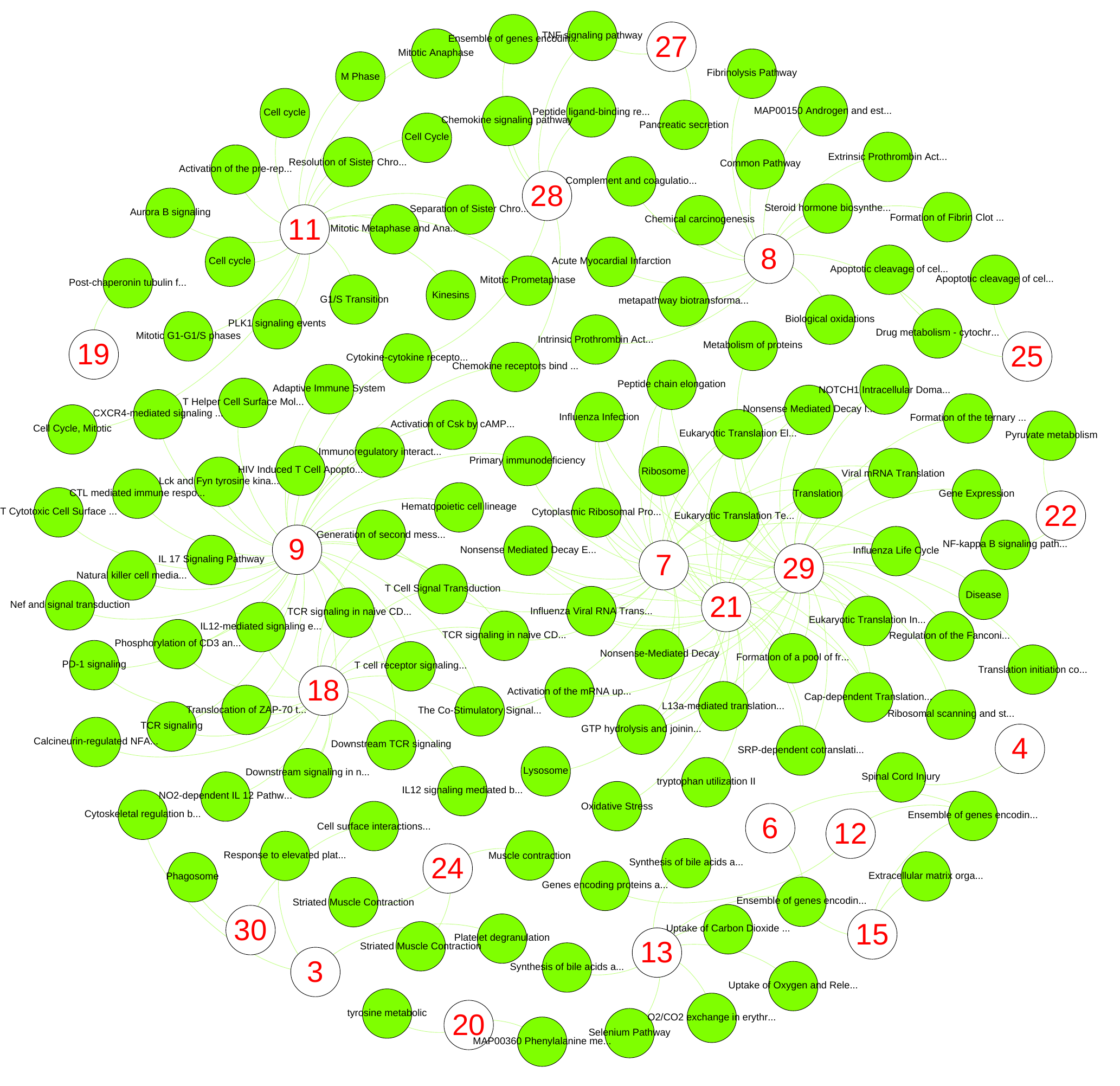}
	\end{center}
	\caption{}
	\label{fig:pathways}
\end{figure}

To analyze the pathway and Gene Ontology (GO) of these 30 gene sets, we used ToppCluster, an online tool for  enrichment analysis of multiple gene sets \cite{Kaimal:2010bv}. The gene set enrichment $p$-values were adjusted using the Bonferroni method. For each gene set, all biological processes, cellular compartments, and pathways that were significantly enriched were analyzed; see Fig. \ref{fig:cc}  for cellular compartments, Fig. \ref{fig:bp} for biological processes, and Fig. \ref{fig:pathways} for pathways. A value of 0.05 was used as the cutoff for the adjusted $p$-values of pathways. To have fewer nodes for better visualization,  we used a slightly more stringent cutoff of 0.01 in cellular compartments and biological processes analyses.

For instance, the CIC component 11 (the lower-left side in Fig. \ref{fig:cc})  is linked to cellular compartments such as spindle, kinetochore, condensed chromosome, centromere, and microtubule. In Fig. \ref{fig:bp} lower-right side, the same component is connected to biological processes such as mitotic cell cycle, mitotic nuclear division, metaphase/anaphase transition, chromosome segregation, nuclear division, and cell division. The same component is connected to cell cycle, M-phase, resolution of sister chromatids, separation of sister chromatids, G1/S transition, mitotic G1-G1/S phases, and cell-cycle pathways; see Fig. \ref{fig:pathways} upper-left side. It is evident that the CIC component 11 represents the genes, pathways,  compartments, and processes involved in cell division. 

While there is an obvious consistency in pathways, compartments, and processes that are linked to cell identity code components, each component is connected to different areas of cellular life. Among cellular compartments, we can see the extracellular space (CIC component 6), lateral plasma membrane (CIC component 25), cytosolic part (CIC component 7), ribosome (CIC component 21), and the nucleolus (CIC component 29). There are a few cases that two or more CIC components are linked to the same compartment (e.g. components 4, 12, 14, 15 and 27 to extracellular space). 

Vast collections of biological processes and pathways are also represented by different CIC components. Likewise, most of them are linked to one CIC component. This shows important aspects of cellular biology are learned by the classifier autoencoders in an unsupervised approach (i.e. without providing any training data about GO or pathways). Our enrichment analysis with ToppCluster showed significant connections between CIC components and additional areas of cell biology such as molecular functions, protein domains, microRNAs, human phenotypes, disorders, and drugs. See supplementary Fig. S1-S6 and Table S1. The outcome of these analyses seem biologically relevant; for instance, CIC component 18 is associated with immunological synapse in Fig. \ref{fig:cc}, T-cell aggregation in Fig. \ref{fig:bp}, T-cell receptor signaling in Fig. \ref{fig:pathways}, and a group of immunity disorders such as Lupus Erythematosus and AIDS; see supplementary Fig. S5.

The stringent choice of Bonferroni $p$-value adjustment and $0.01$ as the cutoff was to reduce the number of nodes for visualization. Using Benjamini-Hochberg false discovery rate $p$-value adjustment and a threshold of $0.05$, which are usually used for GO and pathway analysis, significantly increases the number of pathways, processes and functions that are influenced by CIC components (2115 biological processes, 303 cellular compartments and 484 pathways).  Additional networks and the list of  30 gene sets that are influenced the most by CIC components are provided as supplementary information.

\subsection{CIC Can Resist Noise}

We also questioned how resistant can our model be against noise. This noise can come from the measurement methods such as Microarray or RNA-seq, stochasticity of the gene expression process in the cells, environmental changes or other factors.

For that purpose, we first normalized the expression levels of each gene in all samples between 0 and 1, to control the noise level among all the genes. We performed 5-fold cross-validation. In each round, 80\% of randomly selected samples were used as the training set and the remaining 20\% as the test set. We ensured each sample was used in the test dataset exactly once. Before each round of cross-validation, all weights and bias parameters of the network were restarted to random values to ensure the training dataset of the previous round is unseen in the new round. The noise was generated from a Gaussian distribution $\mathcal{N}\sim(\mu=0, \sigma=0.1)$. During the training, the noise was added to the input layer, but the original GEPs without noise were expected to be reproduced in the output layer. We trained the network for 500 rounds, due to the altered structure of training.

The resulting classification accuracies for the test dataset were between 98.6\% to 100\% in five rounds of cross-validation, with an average accuracy of 99.4\% and a standard error of 0.002. These results confirmed the CIC codes can be used to accurately classify the cell types even in presence of some noise.

\section{Methods}\label{sec:methods}

\subsection{Microarray Data}

The preprocessed microarray expression profiles consisting of 20184 unique genes in 1040 biological samples from 16 human tissues or cell types (65 samples per tissue) were obtained from the CellNet package \cite{Cahan:2014fh}. The cell types included Embryonic Stem Cell (ESC), ovary, skin, neuron, hematopoietic stem and progenitor cells (HSPC), macrophage, B-cell, T-cell, endothelial fibroblast, skeletal muscle, heart, kidney, lung, liver, and colon. All of the data were selected from the NCBI Gene Expression Omnibus (GEO) \cite{Barrett:2013kla} and shared a common microarray platform (Affymetrix HG133 plus 2). There were at least 10 different conditions per cell type/tissue, that was essential for generalization of the learning process and to ensure that our network is not learning only a particular cell type of specific laboratory. 

The raw expression profiles were $\log_2$ transformed and quantile normalized before our analysis. Differential expression analysis was performed using the linear regression and Bayesian analysis of the R/Bioconductor package limma \cite{Ritchie:2015fa}.

\subsection{Deep Neural Networks}
There are many dense clusters in gene regulatory networks and the members of each cluster are usually co-expressed, which made us hypothesize that the information of gene expression profiles (GEPs) can be compressed in a significantly lower dimension, such that the whole GEP can be reproduced from the lower dimension data. To test this hypothesis, we decided to employ autoencoders for dimensionality reduction of the GEPs. 

Autoencoders are a class of deep neural networks that have been applied to important applications such as denoising  \cite{Vincent:2008jr} or  dimensionality reduction and have been shown to outperform other dimensionality reduction algorithms such as PCA for particular classes of  data including images \cite{Hinton:2006bg}. Furthermore, autoencoders can extract useful features from the data and significantly reduce the computational cost of the downstream analysis by replacing the whole data with a vector of smaller dimension. 

A simple form of an autoencoder consists of one hidden layer that is fully connected to both input and output layers. The training data is fed to both input and output layers, and the aim is to reduce the dimensionality of the data in the hidden layer. This form of autoencoder has been used to analyze bacteria \cite{Tan:2016dm} and human gene expression data \cite{Tan:2015vg}. The deep architectures of autoencoders  contain multiple layers between input and output. This type of autoencoder has been used to analyze yeast transcriptome profiles \cite{Chen:2016hz}.

\subsection{Comparison of Autoencoder Architectures}\label{sec:architectures}

The first methodological challenge was to design a particular architecture of the deep autoencoders that could accurately reconstruct the gene expression profiles after training. To address this challenge, we created 10 different architectures of autoencoders, each consisting of an encoder and a decoder part that was attached serially (Fig. \ref{fig:ae}a). They contained different sizes of neurons per layer, and various activation functions. 

For this experiment, we selected 1000 genes with the highest variance of expression among all 1040 samples to speed up the training time. The encoder part of each neural network contained 1000 input neurons and  30 code neurons. Symmetrically, the decoder part started with 30 code neurons and ended with 1000 output neurons. 

\begin{figure}[h!]
	\begin{center}
		\includegraphics[width=1\textwidth]{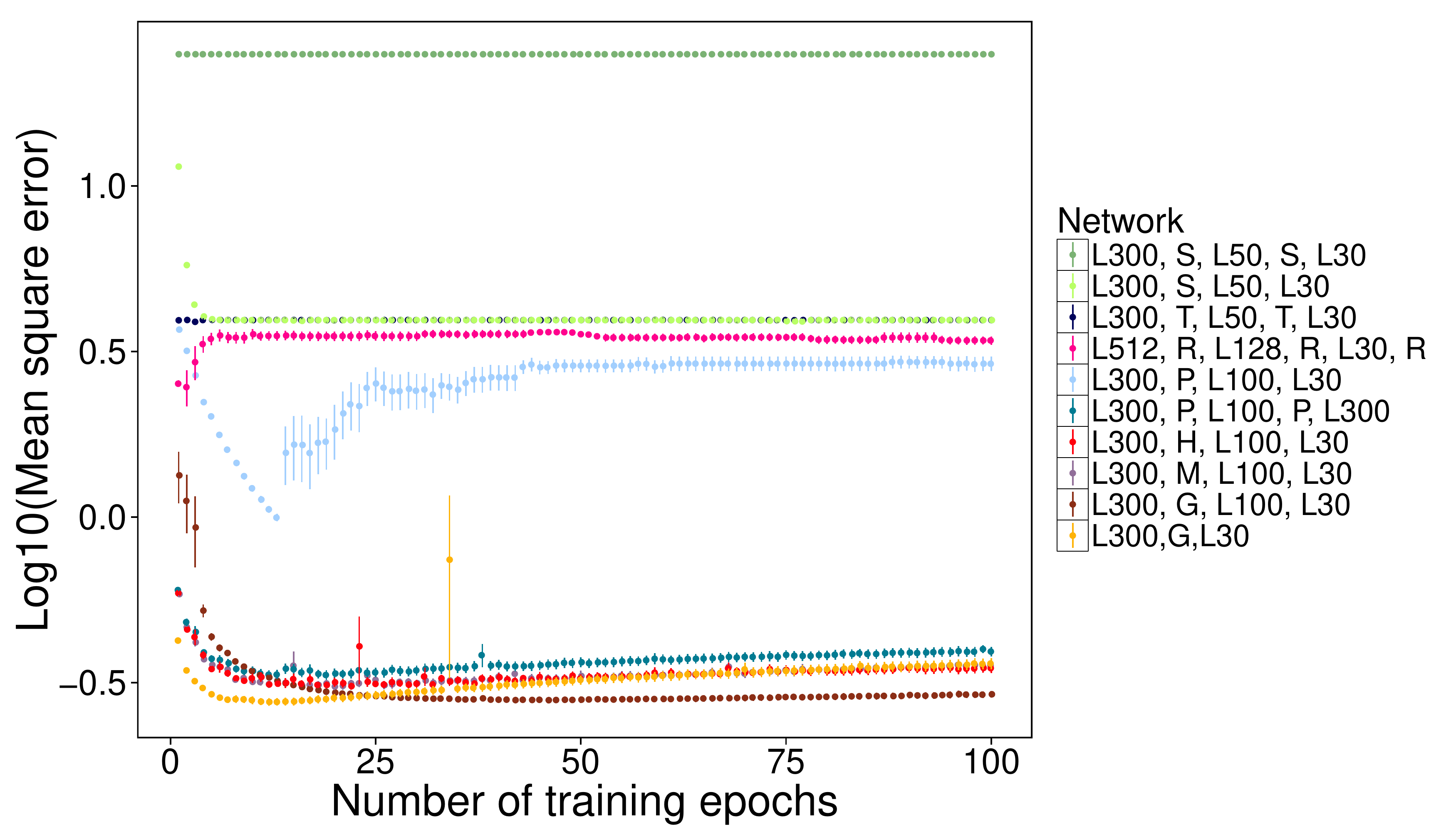}
	\end{center}
	\caption{}
	\label{fig:architecture} 
\end{figure}

The encoder parts included 3 to 5 layers of linear and non-linear neurons; Fig. \ref{fig:architecture}. A subset of 780 samples (75\%) was randomly selected as the training set, and the remaining 260 samples (25\%) were used as the test set. The training phase of each neural network included 100 epochs, with 100 iterations per epoch (total 10,000 iterations). In each iteration, all training samples were fed into the neural network after a random shuffle, and the neural network parameters were updated by a stochastic gradient descent (SGD) algorithm to minimize the mean squared error (MSE) loss function. 

As shown in Fig. \ref{fig:architecture}, five architectures had lower loss values than the others on the test samples. The second to fourth best architectures had a similar 1000 to 300 (1000:300) fully connected linear layer, followed by a Shrink, SoftMax or LogSigmoid function applied to each of the 300 neurons, followed by 300:100 and 100:30 linear layers. The best network had the same architecture as the second best, except that the last two linear layers were merged into a 300:30 fully connected layer. Similar results were expectable from this pair of architectures since the product of the weight matrices for serial linear layers can result in the weight matrix of a single linear layer.

To ensure the stability of the results, we performed 10-fold cross-validation. Error-bars show the standard errors of the MSE values. Standard errors were generally very small, particularly after epoch 40, that showed measured MSEs are stable and the models are not overfitted to a particular portion of the training data.

\subsection{Comparison of Cell Identity Code Dimensionality}\label{sec:dimensions}

The second methodological challenge was to determine the best size of code to reduce the dimensionality of the GEPs without loss of data. For this purpose, we created a set of 10  autoencoders similar to the optimal architecture in the previous experiment, but different sizes of the code layer. In this experiment, we used the complete expression profiles of 20184 genes.  All autoencoders included a 20184:2000 fully connected linear layer,  a LogSigmoid layer, and then the 2000 neurons were fully connected linearly to a code layer of size 10 to 100 neurons. Again, 75\% of data was randomly selected for training, and the remaining 25\% was used as the test data. 

\begin{figure}[h!]
	\begin{center}
		\includegraphics[width=1\textwidth]{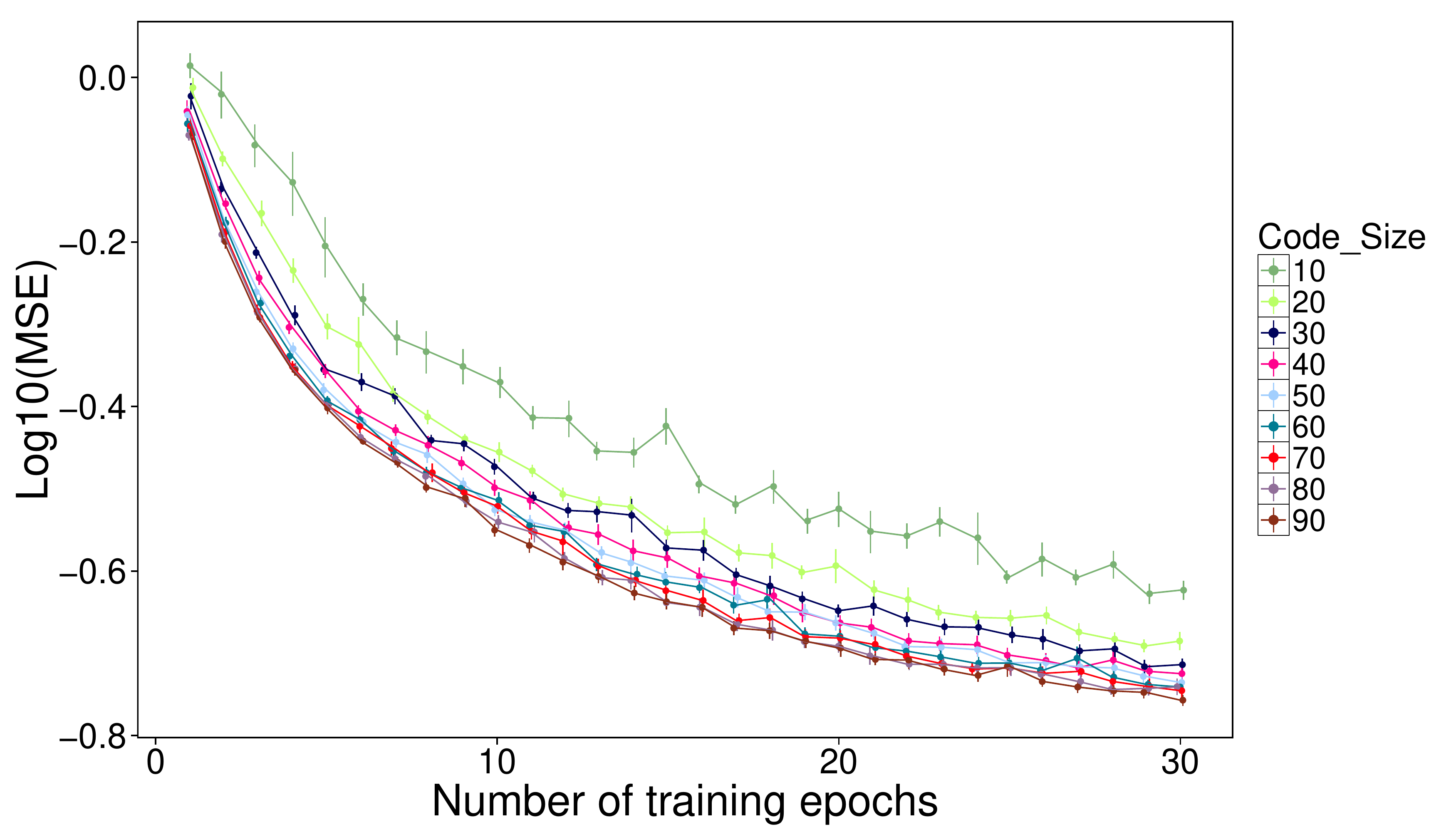}
	\end{center}
	\caption{}
	\label{fig:fpsize} 
\end{figure}

The performance of the networks on the test samples are shown in Fig. \ref{fig:fpsize}. After running 30 epochs of 100 iterations, there is a small gap between the networks with 10 and 20 code-layer neurons and the other networks. In a trade-off between the size of the CIC and the accuracy of the results, we selected 30 as the size of cell identity code for our analyses.

One further observation was the lower MSE levels in this experiment, in comparison with the previous experiment on 1000 genes with the highest expression variances. In the presence of many genes with subtle changes in expression among different samples, a lower error rate was reasonable here.

Again, we performed 10-fold cross-validation without stratification. The standard errors, depicted as the error bars, are quite small that means the robustness of our results.

\subsection{Architecture, training and testing the networks}

We trained an autoencoder with the optimal architecture that we determined previously to compress the gene expression profiles of 20184 genes in 780 training samples for 100 epochs and 100 iterations per epoch, with a total of 10\,000 iterations. In each iteration, one randomly selected training sample was fed into the neural network, and all of the weights were updated by an SGD algorithm. The learning rate was reduced during iterations of the same epoch, but restarted to the original value by each new epoch. Training for 200 epochs resulted in overfitting, see supplementary Fig. S8.

We used the sequential network architecture in all of our scenarios. Training was performed using the stochastic gradient descent algorithm with mean squared error (MSE) criterion. The learning rate of 0.01 did not work very well for many cases; hence, we used the learning rates between 0.001 and 0.003. There were a number of epochs in each training procedure. Each epoch consisted of several iterations, and training data were fed into the network in each iteration. The network parameters were saved after each epoch and used for the next epoch, but the decayed learning rate restarted after each epoch. For many cases, we used 100 epochs and 100 iterations per epoch.

\paragraph*{Transfer and criterion functions.}
To have non-linear layers, we used several typical transfer functions mentioned above. The formulas for the functions are as follows:
$$
ReLU(x) = \max(0,x)\qquad
Sigmoid(x) = \frac{1}{1 + e^{-x}}\qquad
LogSigmoid(x) = \log(\frac{1}{1 + e^{-x}})
$$

$$
Tanh(x) = \frac{e^x - e^{-x}}{e^x + e^{-x}}\qquad
SoftPlus(x) = \log(1 + e^{x})\qquad
SoftShrink(x) = \begin{cases}
x-\lambda, & \text{if}\ x > \lambda\\
x+\lambda, & \text{if}\ x < -\lambda\\
0, & \text{otherwise}
\end{cases}
$$
The soft-max function is applied to a list of values and returns a list of the same size with values in the range $[0,1]$ summing to one, therefore resembling a discrete probability distribution:
$$
  SoftMax_i(x) = \frac{e^{x_i}}{\sum_j{e^{x_j}}}
$$
We also used cross-entropy error as the criterion for the classifier. Assuming we know the correct class $c$ (e.g. cell-type), the cross-entropy loss can be calculated on the output of a soft-max function as follows:
$$
  CE\_Loss(x, c) = -\log\Big(\frac{e^{x_{c}}}{\sum_j{e^{x_j}}}\Big) = -x_{c} + \log\big(\sum_j{e^{x_j}}\big)
$$

More details about different types of activation functions can be found on  "Activation function" page of Wikipedia.
\subsection{Cross-validation}

During the work, we had to ensure the results were robust and the training was not overfitted towards a particular portion of the data. For this purpose, we performed additional experiments using 10-fold cross-validation. In our normal experiments, we randomly selected 75\% of the samples as the training dataset and the remaining 25\% as the test dataset. In 10-fold cross-validations, however, we randomly partitioned all of the samples among 10 groups of equal size. In each round of cross-validation, one group was taken as the test dataset, and the other 9 groups were used as training dataset. Each round of training was started from the scratch, i.e. the network parameters such as weights and biases were restarted to the random initial values. By this way, we ensured the test group is unseen after training the network with the other 9 groups. The test results of all of 10 rounds, such as MSE or correlation values, were merged together by calculating the mean value and standard error. 

\subsection{Gene Set Enrichment Analysis}

We used ToppCluster multi gene-list enrichment analysis online application to determine GO terms, pathways, diseases, drugs, domains, and microRNAs associated with the 30 gene lists associated with the cell identity code components \cite{Kaimal:2010bv}. Nominal $p$-values were adjusted using Bonferroni or Benjamini-Hochberg methods, with 0.01 or 0.05 as $p$-values. While all of the different settings are considered as statistically significant, we increased stringency for some cases to keep the number of nodes suitable for visualization. A complete list of enriched terms can be determined using supplementary Table S1.

\subsection{Visualization}

Both pre- and post-processing of the data and visualization of results were achieved by custom scripts in the R statistical language. We used several R/Bioconductor packages including ggplot2, parallel, data.table, and plyr. Networks were visualized using Gephi \cite{Bastian:2009tf}.

\subsection{Implementation}
We used the script language Lua with the package Torch7 to implement deep neural networks. To increase efficiency, we used Graphical Processing Unit (GPU) through the library CUDA for some of the training procedures. 
In each training procedure, the whole data was read from tabular text files and all samples were permuted using a fixed random seed. A random subset of 75\% of all samples was used to train the networks, and the remaining 25\% to test. The data was transformed into Torch Tensor for CPU, and Cuda Tensor for GPU  training/testing. 

We used several Torch7 packages, including nn, torch, cutorch, cunn and cudnn. Several neuron types were used in our analyses including the linear fully connected layers, rectified linear units (ReLUs), sigmoid, logarithmic sigmoid, hyperbolic tangent, soft-max, soft-plus, and shrink.

We also developed the same architectures in TensorFlow, and did not observe a change in the results due to platform change.

\subsection{Hardware}

We used a Linux server running Fedora 24 version 4.7.5-200. It contained 4 AMD Opteron(tm)  6386 SE processors, with 64 total cores running at 2.8 GHz and 512 GBytes of main memory. Training of neural networks was performed using the Cuda driver on an NVIDIA Tesla K20c GPU running at 706 MHz with 5120 MB memory.

\section{Future works}
The present work can be extended in several ways.
Batch normalization can be employed to limit the variation of values in different genes and datasets. Weight normalization can help to control the variations of weights in each layer. Moreover, more efficacious activation functions such as ReLu, PeakyReLu, Swish, etc. can be used to avoid some issues such as vanishing moments and saturation which can dramatically decrease the network's performance. Finally, hyperparameter tuning algorithms such as grid or random search can be used to evaluate different networks and chose the best one among all.

\section{Acknowledgments}

Authors would like to acknowledge creative comments and ideas by Dr. Mehdi Totonchi and Dr. S. M. Ali Eslami.

\section{Author Contributions} H.C and A.S.Z conceived the project and S.A.M provided ideas for the methods. F.A, B.A, A.M, A.K, and A.S.Z implemented the project and analyzed the data. A.S.Z, H.C, and A.K wrote the manuscript.

\section{Competing Interests} The authors declare no competing interests.

\subsection{Data Availability}

The full source codes and pre-processed data are available upon request.

\subsection{Supplementary Information}
\subsubsection{Supplementary Figure S1} Human phenotypes associated with CIC components.
\subsubsection{Supplementary Figure S2} GO Molecular Functions associated with CIC components.
\subsubsection{Supplementary Figure S3} Protein Domains associated with CIC components.
\subsubsection{Supplementary Figure S4} MicroRNAs associated with CIC components.
\subsubsection{Supplementary Figure S5} Diseases associated with CIC components.
\subsubsection{Supplementary Figure S6} Drugs associated with CIC components.
\subsubsection{Supplementary Figure S7} Pseudocode and the results of the comparison of Cell Identity Code with PCA.
\subsubsection{Supplementary Figure S8} Training the network for 200 epochs results in overfitting.
\subsubsection{Supplementary Table S1} Top 100 genes influenced by each CIC component.

\bibliography{papers}
\pagebreak
\section{Figure Captions}

Figure \ref{fig:ae}: The architecture of autoencoder. (a) We call the output of the red layer in the middle Cell Identity Code (CIC). Each layer is fully connected to the next layer, and the edges are not shown. (b) The network that outperformed the other networks of our study. The number of neurons in each layer and the activation functions are shown above each layer.
\\
\\
Figure \ref{fig:allgenes}: Comparison of the reproduced gene expression profiles (REPs) that are computationally made from cell identity codes (CICs) and the original gene expression profiles (GEPs). (a) The measured mean squared error (MSE) over the test samples during 100 epochs of training. Each dot represents the mean value for 10 measurements during 10-fold cross-validation, and error bars show standard errors. (b) The ranking of similarity of each REP to the original GEP, in comparison with all other test-cases. The similarity is measured by Spearman correlation coefficient. Rank 1 shows the number of test cases that their REPs have been the most similar to their GEPs than any other test-case GEPs. (c) A similar ranking, but the similarity is measured by the MSE value, the lower the MSE value the higher the similarity. (d) Three examples of test-cases from Colon (left column), Neuron (middle column), and ESC (right column). In each column, the first row (red scatter-plots) compare two GEPs of very similar biological samples obtained from the same experiment. The second row shows one of those GEPs, along with its REP. (e) Boxplots that show the quantiles and median of the distribution of Spearman correlation coefficient between GEPs of each test sample and the closest test sample to it (GEP-GEP), and also between each test sample GEP and its REP. (f) The same comparison for MSE in $\log_{10}$ scale. (g) Comparison of the MSE distance between original GEP and reproduced expression profile by using different algorithms. MSE=Mean squared error, COR=Spearman correlation coefficient, GEP=Gene expression profile, REP=Reproduced gene expression profile, CIC=Cell identity code, ESC=Embryonic stem cells, PCA=Principal components analysis, NMF=Non-negative matrix factorization, RBF=Radial basis function, Cos=Cosine.
\\
\\
Figure \ref{fig:generalize}: Cell identity codes can be generated for the cell types and tissues that are not used in training. The median of MSE and Spearman correlation coefficients are provided for test results. See text for more details. 
\\
\\
Figure \ref{fig:classifier}: The schematic architecture of a classifier autoencoder. A gene expression profile (GEP) is encoded into a cell identity code (CIC, the red nodes). The CIC is connected to two parts: the decoder part which reproduces the expression profiles (REP) and also the classifier that identifies the cell type.
\\
\\
Figure \ref{fig:cc}: Cellular compartments that are enriched by each cell identity code (CIC) component. The red numbers show the 30 CIC components. For each CIC neuron, we identified a cluster of top 100 genes that their reproduced expression levels had maximum absolute influence after altering the output of the neuron. Then we identified cellular components that are significantly associated with the cluster of each neuron, which are the additional nodes in the plot. Each link in the plot shows a connection between a CIC component and a cellular compartment that is significantly associated with the genes altered by that CIC component.
\\
\\
Figure \ref{fig:bp}: Biological processes that are enriched by each cell identity code (CIC) component. More details are provided in the caption of Fig. \ref{fig:cc}.  
\\
\\
Figure \ref{fig:pathways}: Pathways that are enriched by each cell identity code (CIC) component. More details are provided in the caption of Fig. \ref{fig:cc}. 
\\
\\
Figure \ref{fig:architecture}: Comparison of 10 different architectures of the deep neural networks in encoding the expression levels of 1000 genes into the cell identity code (CIC) of size 30 and decoding it back to the original values. The numbers following letter L determine the number of neurons after a linear layer, e.g. ``L300, G, L30'' is a linear layer that fully connects the input 1000 neurons to 300 neurons, followed by a layer that applies a LogSigmoid function to each of the 300 neurons, and followed by a second linear layer that fully connects them to 30 neurons of the CIC. The structure of decoder parts are not shown in this figure but are symmetric to the encoders. Each dot indicates the average MSE values (log10 scale) for 10-fold cross-validation, and the error bars show standard errors. 
\\
\\
Figure \ref{fig:fpsize}: Similar structures of autoencoders, having different cell identity code (CIC) sizes are compared in learning the gene expression profiles (GEPs). Each dot indicates the average MSE values (log10 scale) for 10-fold cross-validation, and the error bars show standard errors. 
\pagebreak
\section{Figures}

\end{document}